\documentclass[conference]{IEEEtran}
\IEEEoverridecommandlockouts
\usepackage[american]{babel}
\usepackage[autostyle]{csquotes}
\usepackage{cite}
\usepackage{amsmath,amssymb,amsfonts}
\usepackage{algorithmic}
\usepackage{graphicx}
\usepackage{array}
\usepackage{multirow}
\usepackage{booktabs}   
\usepackage{subcaption}  
\usepackage[colorlinks=true, linkcolor=black, citecolor=black, urlcolor=black]{hyperref}
\usepackage{textcomp}
\usepackage{xcolor}
\def\BibTeX{{\rm B\kern-.05em{\sc i\kern-.025em b}\kern-.08em
    T\kern-.1667em\lower.7ex\hbox{E}\kern-.125emX}}
\begin{document}

\title{Audio-Guided Visual Perception for Audio-Visual Navigation
}

\author{
\parbox{1.0\linewidth}{
\centering
Yi Wang$^{1,2}$, Yinfeng Yu$^{1,2,\dagger}$, Fuchun Sun$^{3}$, Liejun Wang$^{1,2}$, and Wendong Zheng$^{4}$%
\thanks{$^{\dagger}$Yinfeng Yu is the corresponding author (Email: yuyinfeng@xju.edu.cn).}
\\[0.5em] 
\small
\textsuperscript{1}School of Computer Science and Technology, Xinjiang University, Urumqi, China \\
\textsuperscript{2}Joint International Research Laboratory of Silk Road Multilingual Cognitive Computing \\
\textsuperscript{3}Tsinghua University, Beijing, China \\
\textsuperscript{4}Tianjin University of Technology, Tianjin, China
}
}

\maketitle

\begin{abstract}
Audio-Visual Embodied Navigation aims to enable agents to autonomously navigate to sound sources in unknown 3D environments using auditory cues. While current AVN methods excel on in-distribution sound sources, they exhibit poor cross-source generalization: navigation success rates plummet and search paths become excessively long when agents encounter unheard sounds or unseen environments. This limitation stems from the lack of explicit alignment mechanisms between auditory signals and corresponding visual regions. Policies tend to memorize spurious \enquote{acoustic fingerprint-scenario} correlations during training, leading to blind exploration when exposed to novel sound sources. To address this, we propose the AGVP framework, which transforms sound from policy-memorable acoustic fingerprint cues into spatial guidance. The framework first extracts global auditory context via audio self-attention, then uses this context as queries to guide visual feature attention, highlighting sound-source-related regions at the feature level. Subsequent temporal modeling and policy optimization are then performed. This design, centered on interpretable cross-modal alignment and region reweighting, reduces dependency on specific acoustic fingerprints. Experimental results demonstrate that AGVP improves both navigation efficiency and robustness while achieving superior cross-scenario generalization on previously unheard sounds.
\end{abstract}


\section{Introduction}
Moving through the physical world is a naturally multimodal sensory experience: we not only rely on vision to navigate but also use audition to detect distant cues like doorbells, infants crying, or kettles boiling. Audio-visual embodied agents perceive both visual and auditory modalities through sensors, enabling autonomous localization and navigation to sound sources in real, uncharted 3D environments~\cite{soundspaces,ss-001}. For heard sound sources (those seen during training), state-of-the-art AVN methods achieve success rates (SR) exceeding 95\%, nearing the task ceiling. However, future agents must handle diverse, open-world sound sources-implying that we cannot pre-train agents on all possible sounds. When deployed with unheard sounds or unseen environments, current AVN performance plummets, exposing critical generalization bottlenecks across sound sources and scenarios. We attribute this to two core limitations: 1)Mainstream AVN methods concatenate visual and auditory features at the policy level, lacking explicit alignment between sound-source-related visual regions and auditory signals. This prevents sound from guiding vision to \enquote{where to look} when targets are occluded or invisible. 2)The current policy network tends to memorize spurious \enquote{acoustic fingerprint-scenario} correlations during training, leading to excellent navigation performance on heard sound sources. However, this memory-based mapping fails to generalize when encountering unheard sounds.

\begin{figure*}
    \centering
    \includegraphics[width=1\linewidth]{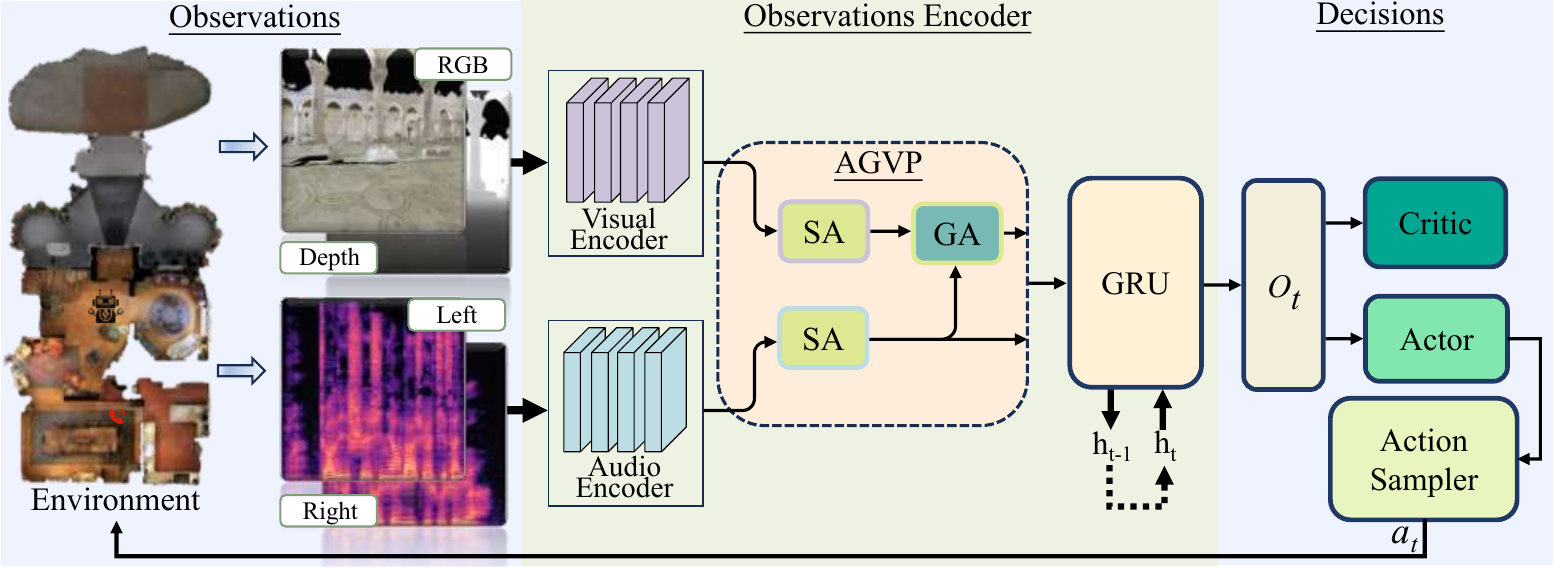}
    \caption{Audio-guided visual perception framework for audio-visual navigation.}
    \label{fig1}
\end{figure*}

In a sudden fire at night, smoke obscures the camera's vision in an apartment; faint cries for help emerge from a distant emergency exit. With visibility blocked, today's embodied agents can only perform random exploration like headless flies in thick smoke, facing collision risks with walls or obstacles and the likelihood of missing the exit. By contrast, a normal-hearing human first turns toward the sound, quickly locks onto a coarse direction, focuses attention on the region where the source is likely to appear, and then uses residual vision to confirm and act. 
Inspired by this, we propose an Audio-Guided Visual Perception (AGVP) framework that turns sound from a policy-memorized acoustic signature into a spatial pointer. Before any heavy visual reasoning, AGVP uses the audio context to recalibrate the visual feature map, highlighting regions most relevant to the source. In this way, audio decides where to look while vision refines how to look; the policy no longer depends on acoustic-signature memory and can promptly orient to unheard sounds and move efficiently toward the target.

In summary, our main contributions are as follows:
\begin{itemize}
\item Late-stage fusion results in poor audio-visual alignment and weak cross-modal association, leading to policy degradation into highly redundant exploration when targets are occluded or invisible;Policies over-rely on memorizing \enquote{acoustic fingerprint-scenario} correlations, exhibiting weaker generalization capability when encountering unheard sounds.

\item We propose an Audio-Guided Visual Perception (AGVP) framework for audio-visual navigation, and design a \enquote{sound first, vision follows} multimodal fusion mechanism for audio-visual navigation.

\item We conduct comprehensive experiments in 3D environments and compare our results with multiple state-of-the-art baseline methods, demonstrating the effectiveness of our approach.
\end{itemize}

\section{Related Work}
SoundSpaces~\cite{soundspaces} represents the pioneering work in establishing the first publicly available simulation platform for Audio-Visual Embodied Navigation~\cite{ss-001}. This foundational work introduced benchmark protocols for audio-visual embodied navigation and released the collected audio datasets, significantly advancing research in this domain. Our work builds upon the SoundSpaces~\cite{soundspaces} platform.

AV-WaN~\cite{Waypoint} effectively addresses the challenge of prolonged exploration in large-scale 3D environments by introducing a waypoint-based navigation strategy to facilitate target sound source discovery. SAVi~\cite{SAVi} enables continuous navigation after sound cessation through a pre-trained sub-module that predicts target sound semantics. CMHM~\cite{catchme} achieves tracking of slowly moving sound sources by learning associations between local maps and auditory signals. SAAVN~\cite{SAAVN} constructs a complex acoustic environment where agents compete with sound attackers under interference, enhancing robustness transferability to clean environments or scenarios with malicious attackers employing random strategies. FSAAVN~\cite{FSAVVN} proposes a novel audio-visual fusion strategy. ORAN~\cite{ORAN} proposes a cross-task navigation skill transfer method. ENMuS\textsuperscript{3}~\cite{tjavn} proposes a framework for audio-visual navigation in multi-sound-source scenarios. Despite the excellent performance of their work~\cite{ttt,wang2023learning,yu2023measuring}, where state-of-the-art methods achieve over 95\% success rate on Replica~\cite{replica}, the agent's performance drops to 50\% when navigating to previously unheard sounds.

The work most closely related to ours is FSAAVN~\cite{FSAVVN}, which proposes a novel audio-visual fusion module named FSA (Feature Self-Attention). This shares a similar motivation with our approach in cross-modal alignment. However, as FSAAVN~\cite{FSAVVN} was designed and evaluated for moving sound source tracking tasks, while our work focuses on static sound source navigation, the experimental setups are not directly comparable. We plan to investigate the transfer of FSAAVN's~\cite{FSAVVN} fusion layer to static scenarios for comparative validation in future work.

\section{Audio Guided Visual Perception}

\begin{figure}[htbp]
  \centering
  \begin{subfigure}[b]{0.48\columnwidth}
    \centering
    \includegraphics[width=\linewidth]{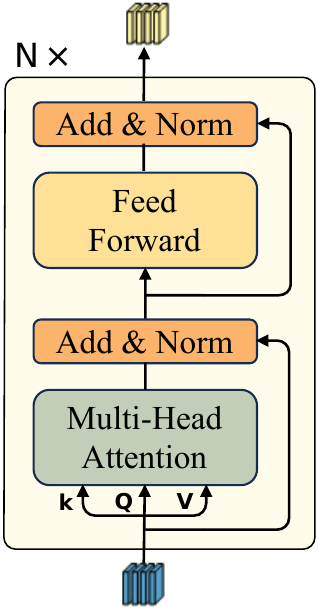}
    \caption{Self-Attention(SA)}  
    \label{fig2.1}
  \end{subfigure}
  \hfill
  \begin{subfigure}[b]{0.48\columnwidth}
    \centering
    \includegraphics[width=\linewidth]{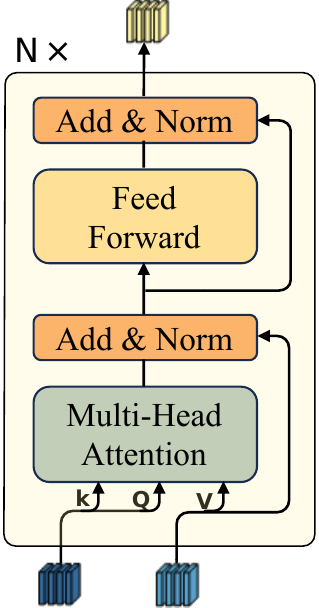}
    \caption{Guided-Attention(GA)}  
    \label{fig2.2}
  \end{subfigure}
  \caption{Core components of AGVP.}  
  \label{fig:2}
\end{figure}

As illustrated in the Fig.~\ref{fig1}, our proposed end-to-end Audio-Guided Visual Perception (AGVP) framework consists of three stages: observation, observation encoding, and policy update based on Proximal Policy Optimization (PPO). The agent continuously explores the environment, acquiring visual and auditory sensory inputs through sensors, which are encoded into features by their respective encoders. Specifically, visual inputs include Depth maps or RGB images, while auditory inputs consist of binaural spectrograms~\cite{binaural}. During the observation encoding stage, the audio and visual features are jointly fed into our AGVP module. First, self-attention~\cite{attention} is applied to the audio sequence to construct a global context. This context then serves as the Query to perform cross-modal guided attention on the visual feature map, explicitly aligning and amplifying the visual regions most relevant to the sound source at the feature level. The aligned multi-modal representation is subsequently passed to a Gated Recurrent Unit (GRU) for temporal modeling across time steps. Finally, the PPO based Actor–Critic head generates action distributions and state values based on the GRU hidden states, completing the closed-loop from \enquote{audio-first visual-following} multi-modal perception to decision-making. Our AGVP module is primarily composed (Fig.~\ref{fig:2}) of Transformer~\cite{attention} based Self-Attention (SA)~\cite{attention,mcan} and Guided-Attention (GA)~\cite{attention,mcan}, whose internal architectures and operational mechanisms will be detailed in the following sections.

\subsection{Self-Attention}\label{AA}
In our framework, the Self-Attention (SA) (Fig.~\ref{fig2.1})~\cite{mcan,attention} module processes the audio and visual modalities independently to better support subsequent cross-modal. When the input is the audio feature $A\in \mathbb{R}^{B\times T\times d}$, the SA module employs a multi-head self-attention mechanism to capture long-range dependencies between frames within the audio sequence, forming an audio representation with global temporal context and producing the enhanced audio representation $A_{\mathrm{SA}} \in \mathbb{R}^{B\times T\times d}$; when the input is the visual feature $V\in \mathbb{R}^{B\times N\times d}$, the SA module computes region-to-region correlations among visual regions. Through such inter region self-attention, it captures the spatial structure information of visual data, thereby obtaining a more context-aware visual representation and outputting the enhanced visual representation $V_{\mathrm{SA}} \in \mathbb{R}^{B\times T\times d}$. This design strengthens the intra modal representational capacity of both modalities.

\subsection{Guided-Attention}
As shown in the Fig.~\ref{fig2.2}, the Guided-Attention (GA)~\cite{attention,mcan} is designed to realize cross-modal guidance and alignment from audio to vision. Unlike the SA module in~\ref{AA}, GA takes two feature streams from different modalities as inputs and uses one modality as guidance (Query) to perform attention over the other modality (Key and Value), thereby explicitly capturing cross-modal correlations and enhancing the representational capacity of the target modality.

Within our AGVP framework, GA specifically uses the audio contextual features as the Query and the visual features as the Key and Value, establishing a cross-modal attention mechanism from sound to vision. Concretely, the inputs to GA are the audio context representation $A_{\mathrm{SA}} \in \mathbb{R}^{B\times T\times d}$ and the visual spatial features $V \in \mathbb{R}^{B\times N\times d}$; the output is the audio-guided, enhanced visual representation $V_{\mathrm{GA}} \in \mathbb{R}^{B\times N\times d}$. GA first applies a Multi-Head Guided Attention mechanism, using $A_{\mathrm{SA}}$ as the Query to guide the selection and aggregation of visual features. The computation proceeds as follows:

First, compute the multi-head guided attention:
\begin{equation}
\mathrm{MHGA}(V,A_{\mathrm{SA}})
= \bigl[\mathrm{head}_1;\dots;\mathrm{head}_h\bigr]W^O,
\end{equation}
with each head given by:
\begin{equation}
\mathrm{head}_i
= \mathrm{softmax}\left(\frac{V W_i^Q (A_{\mathrm{SA}} W_i^K)^\top}{\sqrt{d}}\right) A_{\mathrm{SA}} W_i^V,
\end{equation}
where $W_i^Q,W_i^K,W_i^V\in\mathbb{R}^{d\times (d/h)}$ are the projection matrices of the $i$-th attention head, $W^O\in\mathbb{R}^{d\times d}$ is the output projection matrix, and $\sqrt{d}$ is the dot-product scaling factor.
Subsequently, add the MHGA output to the original visual input via a residual connection \cite{connection} followed by layer normalization \cite{layer}:
\begin{equation}
\widetilde V
= \mathrm{LayerNorm}\bigl(V + \mathrm{MHGA}(V,A_{\mathrm{SA}})\bigr).
\end{equation}

Next, apply a feed-forward network (FFN) for further nonlinear transformation:
\begin{equation}
\mathrm{FFN}(Z)
= \mathrm{ReLU}(Z W_1 + b_1)W_2 + b_2,
\end{equation}
and perform another residual connection with layer normalization to obtain the final output:
\begin{equation}
V_{\mathrm{GA}}
= \mathrm{LayerNorm}\bigl(\widetilde V + \mathrm{FFN}(\widetilde V)\bigr).
\end{equation}

Through these operations, GA achieves explicit audio-to-vision guidance and alignment, strengthening the cross-modal association between visual features and the current audio context. This design enables the agent to localize and track sound sources more precisely, significantly improving navigation efficiency and cross-scene generalization.

\section{Experiments}

\begin{table*}[t]
\centering
\caption{Performance comparison with other methods under the Depth setting. SPL, SR, SNA are percentages.}
\label{tab:depth}
\begingroup
\renewcommand{\arraystretch}{1.05}
\setlength{\extrarowheight}{0pt}
\setlength{\tabcolsep}{2.4pt}
\setlength{\abovetopsep}{0.8ex}
\setlength{\aboverulesep}{0.65ex}
\setlength{\belowrulesep}{0.75ex}
\setlength{\belowbottomsep}{0.8ex}
{\fontsize{12pt}{14.4pt}\selectfont
\begin{tabular*}{\textwidth}{@{\extracolsep{\fill}} l|ccc|ccc|ccc|ccc @{}}
\toprule
\addlinespace[1pt]
& \multicolumn{6}{c|}{\textbf{Replica}} 
& \multicolumn{6}{c}{\textbf{Matterport3D}} \\
\cmidrule(l{8pt}r{8pt}){2-7}\cmidrule(l{8pt}r{8pt}){8-13}
\textbf{Method} &
\multicolumn{3}{c|}{\textbf{Heard}} 
& \multicolumn{3}{c|}{\textbf{Unheard sound}} 
& \multicolumn{3}{c|}{\textbf{Heard}} 
& \multicolumn{3}{c}{\textbf{Unheard sound}} \\
\cmidrule(l{8pt}r{8pt}){2-4}\cmidrule(l{8pt}r{8pt}){5-7}\cmidrule(l{8pt}r{8pt}){8-10}\cmidrule(l{8pt}r{8pt}){11-13}
\addlinespace[1pt]
& \textbf{SPL}$\uparrow$ & \textbf{SR}$\uparrow$ & \textbf{SNA}$\uparrow$
& \textbf{SPL}$\uparrow$ & \textbf{SR}$\uparrow$ & \textbf{SNA}$\uparrow$
& \textbf{SPL}$\uparrow$ & \textbf{SR}$\uparrow$ & \textbf{SNA}$\uparrow$
& \textbf{SPL}$\uparrow$ & \textbf{SR}$\uparrow$ & \textbf{SNA}$\uparrow$ \\
\midrule
Random~\cite{Waypoint}         & 4.9  & 18.5 & 1.8  & 4.9  & 18.5 & 1.8  & 2.1  & 9.1  & 0.8  & 2.1  & 9.1  & 0.8 \\
Direction Follower~\cite{Waypoint}   & 54.7 & 72.0 & 41.1 & 11.1 & 17.2 & 8.4  & 32.3 & 41.2 & 23.8 & 13.9 & 18.0 & 10.7 \\
Frontier Waypoints~\cite{Waypoint}   & 44.0 & 63.9 & 35.2 & 6.5  & 14.8 & 5.1  & 30.6 & 42.8 & 22.2 & 10.9 & 16.4 & 8.1 \\
Supervised Waypoints~\cite{Waypoint} & 59.1 & 88.1 & 48.5 & 14.1 & 43.1 & 10.1 & 21.0 & 36.2 & 16.2 & 4.1  & 8.8  & 2.9 \\
Gan et al.~\cite{gan2020look}           & 57.6 & 83.1 & 47.9 & 7.5  & 15.7 & 5.7  & 22.8 & 37.9 & 17.1 & 5.0  & 10.2 & 3.6 \\
AV-WaN~\cite{Waypoint}               & 86.6 & 98.7 & 70.7 & 34.7 & 52.8 & 27.1 & 72.3 & 93.6 & 54.8 & 40.9 & 56.7 & \textbf{30.6} \\
SoundSpaces~\cite{soundspaces}          & 74.4 & 91.4 & 48.1 & 34.7 & 50.9 & 16.7 & 54.3 & 67.7 & 31.3 & 21.9 & 33.5 & 10.4 \\
\textbf{AGVP (Ours)} & 80.2 & 94.6 & 51.3 & \textbf{46.8} & \textbf{66.5} & \textbf{29.1} & 63.3 & 90.2 & 25.0 & \textbf{41.2} & \textbf{60.8} & 19.1 \\
\addlinespace[1pt]
\bottomrule
\end{tabular*}%
}
\endgroup
\end{table*}

\begin{table*}[t]
\centering
\caption{Performance comparison with other methods under the RGB setting. SPL, SR, SNA are percentages.}
\label{tab:rgb}
\begingroup
\renewcommand{\arraystretch}{0.95}
\setlength{\extrarowheight}{0pt}
\setlength{\tabcolsep}{2.4pt}
\setlength{\abovetopsep}{0.8ex}
\setlength{\aboverulesep}{0.65ex}
\setlength{\belowrulesep}{0.75ex}
\setlength{\belowbottomsep}{0.8ex}
{\fontsize{12pt}{14.4pt}\selectfont
\begin{tabular*}{\textwidth}{@{\extracolsep{\fill}} l|ccc|ccc|ccc|ccc @{}}
\toprule
\addlinespace[1pt]
& \multicolumn{6}{c|}{\textbf{Replica}} 
& \multicolumn{6}{c}{\textbf{Matterport3D}} \\
\cmidrule(l{8pt}r{8pt}){2-7}\cmidrule(l{8pt}r{8pt}){8-13}
\textbf{Method} &
\multicolumn{3}{c|}{\textbf{Heard}} 
& \multicolumn{3}{c|}{\textbf{Unheard sound}} 
& \multicolumn{3}{c|}{\textbf{Heard}} 
& \multicolumn{3}{c}{\textbf{Unheard sound}} \\
\cmidrule(l{8pt}r{8pt}){2-4}\cmidrule(l{8pt}r{8pt}){5-7}\cmidrule(l{8pt}r{8pt}){8-10}\cmidrule(l{8pt}r{8pt}){11-13}
\addlinespace[1pt]
& \textbf{SPL}$\uparrow$ & \textbf{SR}$\uparrow$ & \textbf{SNA}$\uparrow$
& \textbf{SPL}$\uparrow$ & \textbf{SR}$\uparrow$ & \textbf{SNA}$\uparrow$
& \textbf{SPL}$\uparrow$ & \textbf{SR}$\uparrow$ & \textbf{SNA}$\uparrow$
& \textbf{SPL}$\uparrow$ & \textbf{SR}$\uparrow$ & \textbf{SNA}$\uparrow$ \\
\midrule
SoundSpaces~\cite{soundspaces}          & 62.6 & 72.1 & 31.5 & 24.9 & 35.6 & 14.1 & 44.7 & 64.3 & \textbf{22.0} & 20.4 & 30.4 & 7.7 \\
\textbf{AGVP (Ours)} & \textbf{70.3} & \textbf{78.2} & \textbf{37.8} & \textbf{42.5} & \textbf{63.8} & \textbf{16.8} & \textbf{49.8} & \textbf{83.5} & 17.7 & \textbf{21.3} & \textbf{42.0} & \textbf{8.7} \\
\addlinespace[1pt]
\bottomrule
\end{tabular*}%
}
\endgroup
\end{table*}

\subsection{Dataset and Implementation Detail}
We conduct comprehensive experiments using the publicly available Replica \cite{replica} and Matterport3D~\cite{matterport3d} datasets, along with the SoundSpaces~\cite{habitat} acoustic platform built upon these datasets and the Habitat simulator~\cite{habitat}. The Matterport3D~\cite{matterport3d} dataset contains 85 real-world indoor environments primarily focused on residential spaces, providing comprehensive 3D mesh models and corresponding panoramic image scans. The Replica~\cite{replica} dataset comprises 18 synthetic indoor scenes including apartments and hotels, featuring detailed 3D meshes and high-resolution textures, with superior geometric details and photorealistic lighting compared to Matterport3D~\cite{matterport3d}. The sound sources in SoundSpaces are generated by convolving selected audio clips with binaural room impulse responses (RIRs) in corresponding spatial directions, utilizing 102 distinct natural sound categories without replication~\cite{soundspaces}.

In our experiments, we define two distinct sound source conditions:(1) Heard: The target sound source is a telephone ring, used consistently across training, validation, and test splits;(2) Unheard sounds: 102 sound sources split 73/11/18 for train/val/test; evaluation is conducted only on unseen scenes.~\cite{soundspaces}.

\subsection{Evaluation Metrics}
In the experiments, we adopt two metrics to evaluate and compare the navigation performance of different methods: (1) Success weighted by Path Length (SPL): A metric that measures the efficiency of the agent's trajectory by comparing the length of the actual path taken to the optimal shortest path, considering only successful navigation trials~\cite{SPL};(2) Success Rate (SR) : The proportion of test episodes in which the agent successfully reaches the target location;(3) Success-Navigation Accuracy(SNA):Measures the agent's ability to successfully reach the target while maintaining correct orientation during navigation.

\begin{figure*}[t]
    \centering
    \includegraphics[width=1\linewidth]{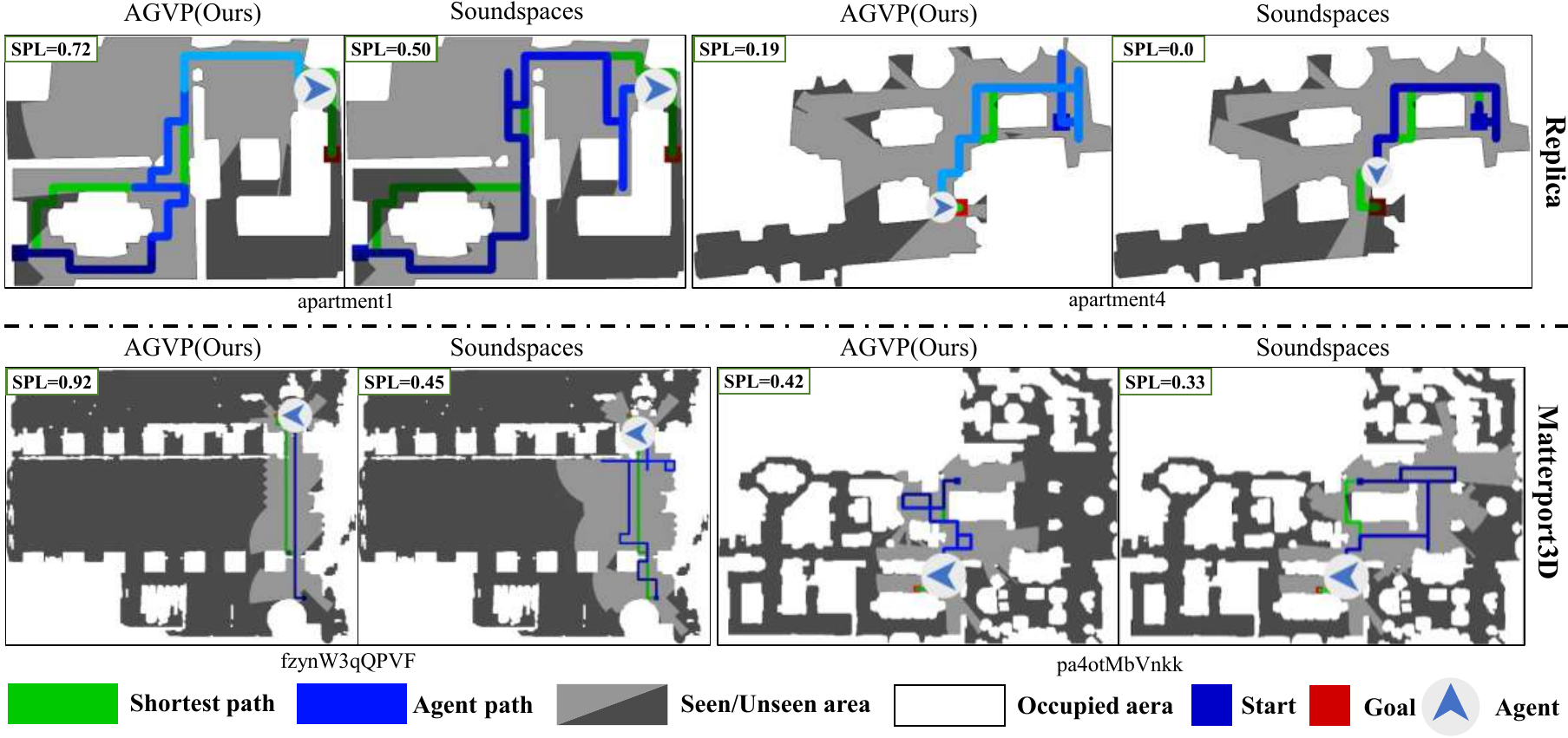}
    
    {\captionsetup{justification=raggedright, singlelinecheck=false}
    \caption{Navigation trajectories on top-down maps.~Agent paths transition from dark to light blue temporally, while green indicates the shortest geodesic path. $\text{SPL} = 0$ indicates that the agent failed to navigate to the target within the prescribed number of steps.}
    \label{fig3}
    }
\end{figure*}

\subsection{Quantitative Experimental Results}

We follow the task setup and evaluation protocol of SoundSpaces~\cite{soundspaces}, with detailed results presented in Table~\ref{tab:depth}. On the Replica dataset, AGVP consistently outperforms the SoundSpaces~\cite{soundspaces} baseline across all sound conditions. In the most challenging \emph{Unheard sound} task, AGVP surpasses AV-WaN~\cite{Waypoint} across all metrics. Specifically, under the \emph{Depth} setting in the \emph{Unheard sound} task, AGVP achieves relative improvements of 34.8\% in SPL, 25.9\% in SR, and 7.4\% in SNA over AV-WaN~\cite{Waypoint}. Even when navigating to completely unseen sound sources, AGVP achieves a success rate (SR) of 66.5\% with significantly higher SPL, highlighting its simultaneous improvement in cross-scenario generalization and path efficiency. When switching to the \emph{RGB} setting (Fig. \ref{tab:rgb})-a more challenging perceptual condition-AGVP maintains strong performance, achieving a success rate of 63.8\%, representing a 79.2\% relative gain over SoundSpaces~\cite{soundspaces}, with SPL and SNA improved by 70.7\% and 19.1\%, respectively, demonstrating strong cross-modal transferability.

On the more challenging Matterport3D~\cite{matterport3d} dataset, AGVP remains consistently superior to the SoundSpaces~\cite{soundspaces} baseline across all sound conditions. In the \emph{Unheard sound} task under the \emph{Depth} setting, AGVP achieves a success rate of 60.8\%, a 7.2\% relative improvement over AV-WaN~\cite{Waypoint}, while also exhibiting better path efficiency. Under the \emph{RGB} setting (Fig. \ref{tab:rgb}), AGVP achieves a 42.0\% success rate when navigating to unseen sound sources, a 38.1\% relative improvement over SoundSpaces~\cite{soundspaces}.

\subsection{Qualitative Experimental Results}
Fig.~\ref{fig3} compares the top-down trajectories of AGVP and the SoundSpaces baseline in 3D environments. It is evident that AGVP generates paths closer to the shortest route while significantly reducing backtracking and wandering, demonstrated by higher SPL scores and shorter effective path lengths. In the scene with ID \textit{fzynW3qQPVF}, AGVP nearly follows the shortest path to the target. This indicates that in scenarios without substantial acoustic occlusion (e.g., no continuous walls or thick obstacles), the audio-guided explicit alignment mechanism at the feature level enables rapid focus on sound-source-related visual regions, clarifying the navigation direction and suppressing inefficient exploration. In contrast, the baseline with only backend concatenation lacks stable audio-visual correspondence, leading to frequent local wandering and repeated backtracking. In the \textit{apartment4} scene, where the sound source is occluded by walls, the SoundSpaces baseline fails to locate the target. By comparison, AGVP still gradually localizes the sound source and successfully completes navigation.

\subsection{Ablation Studies}
To quantify the contribution of each component in the framework, we conduct ablation studies and compare the results with our full model, as shown in Table~\ref{xr}. Removing the cross-modal guidance (w/o GA) leads to a significant performance drop, indicating that aligning audio with visual features and using sound to guide visual processing at the feature level is the key to AGVP's efficiency and stability. Disabling the intra-modal self-attention module (w/o SA) results in degraded performance.

\begin{table}[htbp]
\centering
\caption{Ablation study of AGVP components.}
\label{xr}
\renewcommand{\arraystretch}{1.2}
\setlength{\tabcolsep}{4pt}
\resizebox{\columnwidth}{!}{%
\newcolumntype{C}{>{\large}c}%
\begin{tabular}{l|C|CCC|CCC} 
\toprule
\multirow{2}{*}{\large\textbf{Method}} & \multirow{2}{*}{} 
  & \multicolumn{3}{C|}{\textbf{Replica}}      
  & \multicolumn{3}{C}{\textbf{Matterport3D}} \\
\cmidrule(lr){3-5}\cmidrule(lr){6-8}
& 
& \textbf{SPL} $\uparrow$ & \textbf{SR} $\uparrow$ & \textbf{SNA} $\uparrow$
& \textbf{SPL} $\uparrow$ & \textbf{SR} $\uparrow$ & \textbf{SNA} $\uparrow$ \\
\midrule
\multirow{2}{*}{\large w/o SA}
                     & Depth & 75.4 & 90.6 & 49.0 & 46.6 & 66.1 & 22.1 \\
                     & RGB   & 64.6 & 76.6 & 32.3 & 37.3 & 63.1 & 16.2 \\
\multirow{2}{*}{\large w/o GA}
                     & Depth & 43.1 & 80.1 & 22.4 & 44.9 & 69.6 & 21.5 \\
                     & RGB   & 44.3 & 70.0 & 18.9 & 39.6 & 63.3 & 17.7 \\
\multirow{2}{*}{\large\textbf{AGVP (Ours)}}     
                     & Depth & 80.2 & 94.6 & 51.3 & 63.3 & 90.2 & 25.0 \\
                     & RGB   & 70.3 & 78.2 & 37.8 & 49.8 & 83.5 & 17.7 \\
\bottomrule
\end{tabular}%
}
\end{table}

\section{Conclusions}
We propose the Audio-Guided Visual Perception (AGVP) framework for audio-visual navigation (AVN), advancing multimodal fusion from late-stage, policy-level weighting to the perceptual feature level, where explicit audio-visual alignment enables the principle of \enquote{sound first, vision follows}. AGVP addresses a key limitation of existing AVN methods that rely heavily on feature concatenation: when the target is invisible or occluded, sound alone cannot effectively guide the agent on where to look, forcing the policy network to resort to redundant exploration. In contrast, AGVP leverages auditory context to pre-activate the most sound-source-relevant regions in the visual feature map, which the policy network then uses to complete the closed-loop decision-making process. This mechanism inherently reduces redundant exploration.

AGVP achieves strong cross-scenario generalization on Replica, boosting the success rate to 66.5\%. However, it still has limitations. We envision future work that integrates AGVP with enhanced spatial memory and geometric acoustic modeling, and extends it to multi-source and moving sound source scenarios, to further improve navigation performance on challenging datasets like Matterport3D. We believe this \enquote{listen first, look second, use sound to guide vision} paradigm-based on explicit, feature-level cross-modal alignment-provides a clear starting point for generalizable perception and efficient decision-making in AVN, offering a reusable research framework for future investigations.

\section*{Acknowledgements}

This research was financially supported by the National Natural Science Foundation of China (Grant No. 62463029) and the Natural Science Foundation of Xinjiang Uygur Autonomous Region (Grant No. 2015211C288).

\bibliographystyle{IEEEtran}  
\bibliography{references}     
\end{document}